

\font\twelverm=cmr10 scaled 1200    \font\twelvei=cmmi10 scaled 1200
\font\twelvesy=cmsy10 scaled 1200   \font\twelveex=cmex10 scaled 1200
\font\twelvebf=cmbx10 scaled 1200   \font\twelvesl=cmsl10 scaled 1200
\font\twelvett=cmtt10 scaled 1200   \font\twelveit=cmti10 scaled 1200
\font\twelvesc=cmcsc10 scaled 1200
\skewchar\twelvei='177   \skewchar\twelvesy='60
\def\twelvepoint{\normalbaselineskip=12.4pt
  \abovedisplayskip 12.4pt plus 3pt minus 6pt
  \belowdisplayskip 12.4pt plus 3pt minus 6pt
  \abovedisplayshortskip 0pt plus 3pt
  \belowdisplayshortskip 7.2pt plus 3pt minus 4pt
  \smallskipamount=3.6pt plus1.2pt minus1.2pt
  \medskipamount=7.2pt plus2.4pt minus2.4pt
  \bigskipamount=14.4pt plus4.8pt minus4.8pt
  \def\rm{\fam0\twelverm}          \def\it{\fam\itfam\twelveit}%
  \def\sl{\fam\slfam\twelvesl}     \def\bf{\fam\bffam\twelvebf}%
  \def\mit{\fam 1}                 \def\cal{\fam 2}%
  \def\tt{\twelvett}
  \def\sc{\twelvesc}
  \def\nullspace{\nulldelimiterspace=0pt \mathsurround=0pt }
  \def\big##1{{\hbox{$\left##1\vbox to 10.2pt{}\right.\nullspace$}}}
  \def\Big##1{{\hbox{$\left##1\vbox to 13.8pt{}\right.\nullspace$}}}
  \def\bigg##1{{\hbox{$\left##1\vbox to 17.4pt{}\right.\nullspace$}}}
  \def\Bigg##1{{\hbox{$\left##1\vbox to 21.0pt{}\right.\nullspace$}}}
  \textfont0=\twelverm   \scriptfont0=\tenrm   \scriptscriptfont0=\sevenrm
  \textfont1=\twelvei    \scriptfont1=\teni    \scriptscriptfont1=\seveni
  \textfont2=\twelvesy   \scriptfont2=\tensy   \scriptscriptfont2=\sevensy
  \textfont3=\twelveex   \scriptfont3=\twelveex  \scriptscriptfont3=\twelveex
  \textfont\itfam=\twelveit
  \textfont\slfam=\twelvesl
  \textfont\bffam=\twelvebf \scriptfont\bffam=\tenbf
  \scriptscriptfont\bffam=\sevenbf
  \normalbaselines\rm}

\def\beginlinemode{\endmode
  \begingroup\parskip=0pt \obeylines\def\\{\par}\def\endmode{\par\endgroup}}
\def\beginparmode{\endmode
  \begingroup \def\endmode{\par\endgroup}}
\let\endmode=\par
{\obeylines\gdef\
{}}
\def\singlespace{\baselineskip=\normalbaselineskip}

\def\oneandahalfspace{\baselineskip=\normalbaselineskip
  \multiply\baselineskip by 3 \divide\baselineskip by 2}
\def\doublespace{\baselineskip=\normalbaselineskip \multiply\baselineskip by 2}

\newcount\firstpageno
\firstpageno=2
\footline={\ifnum\pageno<\firstpageno{\hfil}\else{\hfil\twelverm\folio\hfil}\fi}
\let\rawfootnote=\footnote		
\def\footnote#1#2{{\rm\parindent=20pt\singlespace\hang
  \rawfootnote{#1}{\tenrm#2\hfill\vrule
                                  height 0pt depth 6pt width 0pt}}}
\def\raggedcenter{\leftskip=4em plus 12em \rightskip=\leftskip
  \parindent=0pt
  \parfillskip=0pt \spaceskip=.3333em \xspaceskip=.5em
  \pretolerance=9999 \tolerance=9999
  \hyphenpenalty=9999 \exhyphenpenalty=9999 }
\hsize=6.5truein
\vsize=8.9truein
\parskip=\medskipamount
\twelvepoint		
\overfullrule=0pt	
\def\preprintno#1{
 \rightline{\rm #1}}	
\def\title			
  {\null\vskip 3pt plus 0.3fill \beginlinemode
   \doublespace \raggedcenter \bf}
\def\author			
  {\vskip 3pt plus 0.3fill \beginparmode \raggedcenter \sc}
\def\and			
  {\vskip 3pt plus 0.3fill \raggedcenter \rm and}
\def\affil			
  {\vskip 3pt plus 0.1fill \beginlinemode
   \oneandahalfspace \raggedcenter \sl}
\def\abstract			
  {\vskip 3pt plus 0.3fill \beginparmode
   \oneandahalfspace \narrower ABSTRACT:~~}
\def\endtitlepage{\vfill\eject\body}
\def\body{\beginparmode}
\def\subhead#1{\vskip 0.25truein{\raggedcenter #1 \par}
   \nobreak\vskip 0.25truein\nobreak}
\def\references
  {\subhead{REFERENCES}
   \frenchspacing \parindent=0pt \leftskip=0.8truecm \rightskip=0truecm
   \parskip=4pt plus 2pt \everypar{\hangindent=\parindent}}
\def\refstylenp{		
  \gdef\refto##1{~[##1]}				
  \gdef\r##1{~[##1]}	         			
  \gdef\refis##1{\indent\hbox to 0pt{\hss[##1]~}}     	
  \gdef\citerange##1##2##3{~[\cite{##1}--\setbox0=\hbox{\cite{##2}}\cite{##3}]}
  \gdef\journal##1, ##2, ##3,                           
    ##4{{\sl##1} {\bf ##2} (##3) ##4}
  \gdef\eq{eq.}
  \gdef\eqs{eqs.}
  \gdef\Eq{Eq.}
  \gdef\Eqs{Eqs.} }

\def\prd{\journal Phys. Rev. D}
\def\prl{\journal Phys. Rev. Lett.}

\def\pl{\journal Phys. Lett.}

\def\endreferences{\body}

\def\endit{\endmode\vfill\supereject\end}
\def\frac#1#2{{\textstyle{#1 \over #2}}}

\def\sss{\scriptscriptstyle}

\def\twiddle{\lower.9ex\rlap{$\kern-.1em\scriptstyle\sim$}}
\def\bigtwiddle{\lower1.ex\rlap{$\sim$}}
\def\gtwid{\mathrel{\raise.3ex\hbox{$>$\kern-.75em\lower1ex\hbox{$\sim$}}}}
\def\ltwid{\mathrel{\raise.3ex\hbox{$<$\kern-.75em\lower1ex\hbox{$\sim$}}}}
\def\square{\kern1pt\vbox{\hrule height 1.2pt\hbox{\vrule width 1.2pt\hskip 3pt
   \vbox{\vskip 6pt}\hskip 3pt\vrule width 0.6pt}\hrule height 0.6pt}\kern1pt}
\def\ucsb{Department of Physics\\University of California\\
Santa Barbara, CA 93106}

\def\l{\lambda}

\def\r{\rho}

\def\P{\Phi}

\def\tev{{\rm \,Te\kern-0.125em V}}
\def\gev{{\rm \,Ge\kern-0.125em V}}
\def\mev{{\rm \,Me\kern-0.125em V}}
\def\kev{{\rm \,ke\kern-0.125em V}}
\def\ev{{\rm \,e\kern-0.125em V}}

\def\sla{\raise.15ex\hbox{$/$}\kern-.57em}
\refstylenp
\catcode`@=11
\newcount\r@fcount \r@fcount=0
\newcount\r@fcurr
\immediate\newwrite\reffile
\newif\ifr@ffile\r@ffilefalse
\def\w@rnwrite#1{\ifr@ffile\immediate\write\reffile{#1}\fi\message{#1}}
\def\writer@f#1>>{}
\def\referencefile{
  \r@ffiletrue\immediate\openout\reffile=\jobname.ref%
  \def\writer@f##1>>{\ifr@ffile\immediate\write\reffile%
    {\noexpand\refis{##1} = \csname r@fnum##1\endcsname = %
     \expandafter\expandafter\expandafter\strip@t\expandafter%
     \meaning\csname r@ftext\csname r@fnum##1\endcsname\endcsname}\fi}%
  \def\strip@t##1>>{}}

\def\citeall#1{\xdef#1##1{#1{\noexpand\cite{##1}}}}
\def\cite#1{\each@rg\citer@nge{#1}}	
\def\each@rg#1#2{{\let\thecsname=#1\expandafter\first@rg#2,\end,}}
\def\first@rg#1,{\thecsname{#1}\apply@rg}	
\def\apply@rg#1,{\ifx\end#1\let\next=\relax
\else,\thecsname{#1}\let\next=\apply@rg\fi\next}
\def\citer@nge#1{\citedor@nge#1-\end-}	
\def\citer@ngeat#1\end-{#1}
\def\citedor@nge#1-#2-{\ifx\end#2\r@featspace#1 
  \else\citel@@p{#1}{#2}\citer@ngeat\fi}	
\def\citel@@p#1#2{\ifnum#1>#2{\errmessage{Reference range #1-#2\space is bad.}%
    \errhelp{If you cite a series of references by the notation M-N, then M and
    N must be integers, and N must be greater than or equal to M.}}\else%
 {\count0=#1\count1=#2\advance\count1
by1\relax\expandafter\r@fcite\the\count0,%
  \loop\advance\count0 by1\relax
    \ifnum\count0<\count1,\expandafter\r@fcite\the\count0,%
  \repeat}\fi}
\def\r@featspace#1#2 {\r@fcite#1#2,}	
\def\r@fcite#1,{\ifuncit@d{#1}
    \newr@f{#1}%
    \expandafter\gdef\csname r@ftext\number\r@fcount\endcsname%
                     {\message{Reference #1 to be supplied.}%
                      \writer@f#1>>#1 to be supplied.\par}%
 \fi%
 \csname r@fnum#1\endcsname}
\def\ifuncit@d#1{\expandafter\ifx\csname r@fnum#1\endcsname\relax}%
\def\newr@f#1{\global\advance\r@fcount by1%
    \expandafter\xdef\csname r@fnum#1\endcsname{\number\r@fcount}}
\let\r@fis=\refis			
\def\refis#1#2#3\par{\ifuncit@d{#1}
   \newr@f{#1}%
   \w@rnwrite{Reference #1=\number\r@fcount\space is not cited up to now.}\fi%
  \expandafter\gdef\csname r@ftext\csname r@fnum#1\endcsname\endcsname%
  {\writer@f#1>>#2#3\par}}
\def\ignoreuncited{
   \def\refis##1##2##3\par{\ifuncit@d{##1}%
     \else\expandafter\gdef\csname r@ftext\csname
r@fnum##1\endcsname\endcsname%
     {\writer@f##1>>##2##3\par}\fi}}
\def\r@ferr{\endreferences\errmessage{I was expecting to see
\noexpand\endreferences before now;  I have inserted it here.}}
\let\r@ferences=\references
\def\references{\r@ferences\def\endmode{\r@ferr\par\endgroup}}
\let\endr@ferences=\endreferences
\def\endreferences{\r@fcurr=0
  {\loop\ifnum\r@fcurr<\r@fcount
    \advance\r@fcurr by 1\relax\expandafter\r@fis\expandafter{\number\r@fcurr}%
    \csname r@ftext\number\r@fcurr\endcsname%
  \repeat}\gdef\r@ferr{}\endr@ferences}

\def\range#1#2#3{\citerange{#1}{#2}{#3}}
\let\r@fend=\endpaper\gdef\endpaper{\ifr@ffile
\immediate\write16{Cross References written on []\jobname.REF.}\fi\r@fend}
\catcode`@=12
\citeall\refto		
\citeall\r		%

\def\q{\frac14}
\def\db{\Delta\beta}
\def\ku{k_{\rm\sss U}}
\def\mpl{M_{\rm\sss Pl}}
\def\drr{\delta\rho/\rho}
\def\pmax{\P_{\sss\rm MAX}}
\def\psr{\P_{\sss\rm SR}}
\def\pstar{\P_{*}}
\def\tstar{t_{*}}
\def\tc{t_{\rm c}}
\def\pdot{{\dot\P}}
\referencefile
\ignoreuncited
\singlespace
\preprintno{UCSBTH--92--20}
\preprintno{May 1992}
\doublespace
\title
DEPENDENCE OF DENSITY PERTURBATIONS ON THE COUPLING CONSTANT IN A
SIMPLE MODEL OF INFLATION

\author
Toby Falk\footnote{$^{\rm (a)}$}
{Electronic address (internet): falk@tpau.physics.ucsb.edu.}

\author
Raghavan~Rangarajan\footnote{$^{\rm (b)}$}
{Electronic address (internet): raghu@pcs3.ucsb.edu.}

\and

\author
Mark Srednicki\footnote{$^{\rm (c)}$}
{Electronic address (internet): mark@tpau.physics.ucsb.edu.}

\affil\ucsb

\abstract
In the standard inflationary scenario
with inflaton potential $V(\P)=M^4-\frac14\l\P^4$,
the resulting density perturbations $\drr$ are proportional to $\l^{1/2}$.
Upper bounds on $\drr$ require $\l\ltwid10^{-13}$.  Ratra has shown that
an alternative treatment of reheating results in $\drr\propto\l^{-1}$, so that
an upper bound on $\drr$ does not put an obvious upper bound on $\l$.
We verify that $\drr\propto\l^{-1}$ is indeed a possibility, but show
that $\l\ltwid10^{-13}$ is still required.

\endtitlepage
\body
\oneandahalfspace

The inflationary paradigm\range{guth81}{linde82,as82}{linde83}
explains many mysteries of large scale cosmology.
It also provides a source of density fluctuations which act as the seeds
for structure formation, and predicts that these fluctuations have
a Harrison-Zel'dovich spectrum\range{gp82}{hawking82,starobinsky82}{bst83}.
The main problem with the standard inflationary scenario is that it requires
very small self-couplings of the inflaton field $\P$
in order to produce mass fluctuations with the correct amplitude of
$\drr\simeq 10^{-5}$ at horizon crossing.  This is because
$\drr\propto\l^{1/2}$, where $\l$ is the quartic self-coupling of $\P$.
It turns out that $\drr\ltwid 10^{-5}$ requires $\l\ltwid 10^{-13}$.
Many models have been constructed which attempt to make such small couplings
arise naturally.

However, Ratra argues that a very small coupling
may not be necessary\r{ratra91}.
He finds that the dependence of $\drr$ on $\l$ is sensitively dependent
on ``reheating,'' that is, on how the transition from the inflationary era
to the radiation-dominated era is modelled.  In the standard inflationary
scenario, the reheating transition takes place in a few Hubble times.
In Ratra's alternative scenario, reheating is instantaneous (which means,
in practice, much less than a Hubble time).  In this case Ratra finds
that $\drr$ is proportional to $\l^{-1}$, a dramatically different result.
Since, as
Ratra points out, the reheating process is quite complicated, involving
nonequilibrium thermodynamics of a quantum field in curved space,
we should be cautious about adopting a specific model of it unless we
are convinced that its predictions are robust.
It is therefore extremely important to check this point, and to see
whether or not a small $\drr$ can result from a coupling which is
larger than $\lambda\simeq 10^{-13}$.

We have reanalyzed Ratra's results for the simple potential
$$V(\P)=M^4-\q\l\P^4\;,  \eqno(1)$$
where $M$ is a constant, and $\P=0$ at the start of inflation.
Of course, this potential is unbounded below and must be modified for
$\P>\pmax$, where $\pmax=(4/\l)^{1/4}M$ and is defined via
$V(\pmax)=0$.  This potential was originally intended to mock up a
Coleman-Weinberg potential in a gauge theory (in which case $\l\sim g^4$,
where $g$ is the gauge coupling).  This possibility was subsequently
discarded (since $\l\sim g^4$ is much too large), but the prediction
for $\drr$ from the potential of Eq.(1) was thoroughly analyzed in
both the standard scenario and in Ratra's alternative scenario,
and therefore provides a good test case.
Ratra has also analyzed several other possible potentials,
but we will not do so here.  All of our results will apply strictly to
the potential of Eq.(1); we will have nothing to say about Ratra's
other models, although it would be interesting to compare his results
for an exponential potential with those of, for example, Ref.\r{lm85}.

Ratra's analysis includes a complete rederivation of the fluctuation
amplitude and spectrum, making use of gauge noninvariant variables
followed by careful identification of the gauge variant modes.
However, the final result can (necessarily) be derived using the more
standard gauge invariant formalism of Bardeen\r{bardeen80}.
In fact, we can simply
use the final formula of Bardeen, Steinhardt, and Turner (BST)\r{bst83},
without reference to its long derivation.  Many other analyses have confirmed
this formula, except for small differences in the overall normalization.
These will not be relevant, however.

The BST formula for $\drr$ for a perturbation with wavenumber $k$ which
first crossed out of the horizon at time $\tc$ and then reentered during
the matter dominated era is
$${\delta\rho\over\rho}\simeq{1\over5\pi}\,{H^2\over\pdot(\tc)}\;.  \eqno(2)$$
Here $H$ is the Hubble parameter during inflation, related to $M$
via $H=(8\pi/3)^{1/2}M^2\!/\mpl$, where $\mpl$ is the Planck mass.
The field $\P(t)$ is treated as a classical,
spatially uniform, background field;
quantum fluctuations in $\P$ are what ultimately result in the density
fluctuations of Eq.(2).

Clearly, to compute $\drr$ we need to compute $\pdot(\tc)$.
To do so, we use the equation of motion
$$\ddot\P+3H\pdot-\l\P^3=0 \eqno(3)$$
which follows from the potential of Eq.(1).  This equation is easy to solve
in the slow-rollover approximation, where we neglect $\ddot\P$.  When this
approximation is valid we find
$$\P(t)=\left[\pstar^{-2}+\frac23\l H^{-1}(\tstar-t)\right]^{-1/2}
                                                        \;.\eqno(4)$$
Here $\tstar$ is the time when inflation ends, and $\pstar$ is the value
of $\P$ at this time: $\pstar=\P(\tstar)$.  At the moment we will leave
$\pstar$ as a free parameter, but of course we must have
$\pstar\le\pmax$.  The slow-rollover approximation
breaks down when $\ddot\P\simeq3H\pdot$; using Eq.(4), this occurs when
$\P\simeq\psr$, where
$$\psr=\left({3\over\l}\right)^{1/2}H\;.\eqno(5)$$
Thus we must also have $\pstar\le\psr$.
Using Eq.(5), we can rewrite Eq.(4) as
$$\P(t)=\left[\pstar^{-2}+2\psr^{-2}H(\tstar-t)\right]^{-1/2}\;.\eqno(6)$$
Then we can use $3H\pdot=\l\P^3$, valid during the slow-rolling epoch,
to compute $\pdot(\tc)$.
The factor of $H(\tstar-\tc)$ which appears is related to $k$ and $M$ via
$$\db\equiv H(\tstar-\tc)\simeq 69 + \ln(\ku/k) + \ln(M/\mpl)\;,  \eqno(7)$$
where $\ku$ is the wavenumber of the present Hubble radius
($2\pi/\ku\simeq 10^{28}\,{\rm cm}$), and we have implicitly assumed
a reheating temperature of order $M$.  (This is not essential, and was
done only to simplify the formula.)
We ultimately find
$${\delta\rho\over\rho}\simeq{3H^3\over5\pi\l}\left[\pstar^{-2}
                       +2(\db)\psr^{-2}\right]^{3/2} \;. \eqno(8)$$
This is the key equation from which we will be able to understand the
difference between the standard scenario and the alternative scenario.

In the standard scenario, inflation ends when the slow-rollover approximation
breaks down: once $\P$ exceeds $\psr$, the field moves rapidly to
the minimum of the potential.
Thus, in the standard scenario, we have $\pstar\simeq\psr$.  Since $\db\gg 1$,
Eq.(8) implies
$${\delta\rho\over\rho}\simeq{1\over5\pi}\left({8\over3}\right)^{1/2}(\db)^{3/2}
                   \l^{1/2} \qquad\qquad[\hbox{Standard scenario}]. \eqno(9)$$
This is the usual result; in particular, we see that $\drr$ is proportional
to $\l^{1/2}$, and that $\drr\ltwid 10^{-5}$ for $\db\gtwid 45$
requires $\l\ltwid 10^{-13}$.

Ratra, however, suggests that $\pstar$ should not be identified with $\psr$.
Instead, he proposes that $\pstar$ may be much less than $\psr$.  Strictly
within the context of the potential of Eq.(1), this is not possible.
However, we can consider a modified potential, one which drops quickly to
zero for $\P>\pstar$.  In this case, inflation would end when
$\P$ reaches $\pstar$.  This is the scenario that Ratra refers to
as ``rapid reheating.''  If $\pstar \ll (\db)^{-1/2}\psr$, then Eq.(8) yields
$${\delta\rho\over\rho}\simeq{3\over5\pi\l}\left({H\over\pstar}\right)^3
                         \qquad\qquad[\hbox{Alternative scenario}]. \eqno(10)$$
We see that now $\drr$ is proportional to $\l^{-1}$, confirming Ratra's result.

Let us now examine what limits, if any, can be placed on $\l$ in the
alternative scenario.
Since we have a new free parameter, $\pstar$, it would seem that we could
increase $\l$ yet keep $\drr$ fixed by simultaneously decreasing $\pstar$.
This is correct, but only
as long as we remain within the range of validity of Eq.(10),
$\pstar\ll(\db)^{-1/2}\psr$.  From Eq.(5), however, we see that
$\psr$ decreases as $\l$ increases,
so larger values of $\l$ put tighter
constraints on the allowed values of $\pstar$ in the alternative scenario.
To get a global overview, let us start from Eq.(8), which is always valid.
Consider keeping $\lambda$ fixed, and varying $\pstar$ in order
to minimize $\drr$.
It is clear from Eq.(8) that minimizing $\drr$ with $\l$ fixed requires
{\it maximizing} $\pstar$.
But the maximum value of $\pstar$ is
$\psr$, and $\pstar=\psr$ just results in the standard scenario.
This implies that, for a given value of $\lambda$, the smallest possible
$\drr$ is achieved in the standard scenario.
Thus, achieving the same value of $\drr$ in the alternative scenario
requires a {\it smaller} value of $\l$ than is needed in the standard scenario.
For example, to get $\drr\simeq 10^{-5}$ with $\db\simeq 60$
requires $\l\simeq4\times10^{-14}$ in the standard scenario.  In the
alternative scenario with
$\pstar=\frac1{10}(\db)^{-1/2}\psr$, we find that $\l\simeq 3\times 10^{-19}$
is required.  More generally,
it is easy to check that $\pstar=10^{-\nu}(\db)^{-1/2}\psr$
requires $\l\simeq 3\times10^{-13-6\nu}$ for $\nu\gtwid 1$.
Thus we conclude that, while it is possible to arrange a potential
for which $\drr\propto\l^{-1}$, the upper limit on $\l$ actually
{\it decreases}, which is the opposite of the desired goal.

Also, we see that getting
$\drr\propto\l^{-1}$ does not really depend on how much
time it takes for reheating to occur, but rather on when inflation ends.
The important point is whether $\pstar$ is larger or smaller than
$(\db)^{-1/2}\,\psr$.
If $\pstar\gg(\db)^{-1/2}\,\psr$, then inflation ends due to
the increasing acceleration of $\P$ in a smooth potential;
this is the standard scenario.
If $\pstar\ll(\db)^{-1/2}\,\psr$, then inflation ends due to
$\P$ crossing a sudden, sharp feature in the potential;
this is the alternative scenario.
We feel that the two scenarios would be more aptly named ``late turn-off''
and ``early turn-off,'' corresponding to whether inflation ends after or
before $\P$ reaches $(\db)^{-1/2}\,\psr$,
rather than ``slow reheating'' and ``fast reheating.''
As we have seen, whether $\drr$ is proportional to $\l^{1/2}$ or to  $\l^{-1}$
does not actually depend on the speed of reheating,
but rather on the value of the field when inflation ends.

We are very grateful to Bharat Ratra for extensive discussions of his
results.
This work was supported in part by NSF Grant No.~PHY-86-14185.

\vfill\eject
\references

\refis{as82}A. Albrecht and P. Steinhardt, \prl, 48, 1982, 1220.

\refis{bardeen80}J. Bardeen, \prd, 22, 1980, 1882.

\refis{bst83}J. Bardeen, P. Steinhardt, and M. S. Turner, \prd, 28, 1983, 679.

\refis{gp82}A. Guth and S.-Y. Pi, \prl, 49, 1982, 1110.

\refis{guth81}A. Guth, \prd, 23, 1981, 347.

\refis{hawking82}S. W. Hawking, \pl, 115B, 295, 1982.

\refis{linde82}A. D. Linde, \pl, 108B, 1982, 389.  

\refis{linde83}A. D. Linde, \pl, 129B, 1983, 177.  

\refis{lm85}F. Lucchin and S. Matarrese, \prd, 32, 1985, 1316.

\refis{ratra91}B. Ratra, \prd, 44, 1991, 365, and references
to earlier work therein.

\refis{starobinsky82}A. Starobinsky, \pl, 117B, 1982, 175.

\endreferences\endit\end